# Analytical Formulae for Projected Solid Angle on Arbitrary Polygonal Cross Sections

Brett A. Cruden

*Analytical Mechanics Associates Inc. at NASA Ames Research Center, Moffett Field, CA 94035 USA*

Closed form solutions for the computation of the solid angle from polygonal cross-sections are well known, however similar formulae for computation of projected solid angle are not generally available. Formulae for computing the projected solid angle from arbitrarily shaped polygons are derived using the Gauss-Bonnet theorem. This is accomplished by transforming the projected solid angle integral to an integral over a spherical patch, which is then reduced by Gauss-Bonnet to a simple summation over its edges, allowing the projected solid angle to be computed exactly. Application of the formulae allows exact calculation of projected solid angle over discrete intervals which may be used for computing radiative flux to surfaces or view factors to free space.

A common integral in quantitative radiometry is the projected solid angle, $\bar{\Omega}$: [1]

$$\bar{\Omega} = \iint \cos\theta \sin\theta \, d\theta d\varphi \quad (1)$$

The projected solid angle is required for evaluation of irradiance or heat flux at a surface, e.g [2]:

$$E = \iint L \cos\theta \sin\theta \, d\theta d\varphi \quad (2)$$

This quantity is related to the solid angle, $\Omega$

$$\Omega = \iint \sin\theta \, d\theta d\varphi \quad (3)$$

But differs by the $\cos\theta$ factor in the integrand which accounts for the projection of the differential area onto a planar surface. Analytical relationships for computing solid angle of finite shapes are well-known but we are not aware of a similar relation for the projected solid angle. For a discretized radiation field, it is necessary to evaluate the projected solid angle over a discrete grid. In terms of the 2D integral, eq. (1), this means the limits of integration are not necessarily separable in $\theta$ and $\varphi$ such that the analytical solution cannot generally be found by performing integrations of independent variables. Each discrete element of the grid may be described as a planar polygon in space, Fig. 1. The limits of integration over theta and phi are then described by the projection of the polygon onto a unit sphere as shown in Figure 1.

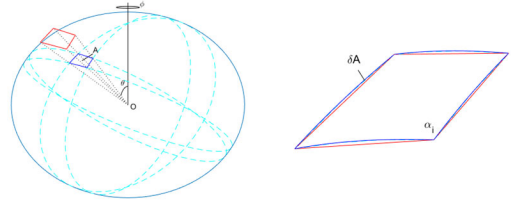

Figure 1. The projected surface on a unit sphere describing the solid angle integral and spherical patch A.

This letter presents an analytical solution to the projected solid angle for a section in space described by an arbitrary polygon. Because the integral is equivalent to evaluating the projection of the polygon onto a sphere, the solution makes use of concepts in non-Euclidean geometry, specifically the Gauss-Bonnet theorem. The Gauss-Bonnet theorem states that the area integral of surface curvature (K) may be related to the sum of the interior angles ($\alpha_i$) of the projected polygon and the integral of the geodesic curvature ($\kappa_g$) over its boundary ($\partial A$):

$$\iint_A K dA = \sum_{i=1}^{N} \alpha_i - (N-2)\pi - \oint_{\partial A} \kappa_g dl \quad (4)$$

Where N is the number of vertices of the polygon. For a spherical surface, the curvature is unity, and the differential area is written as $dA = \sin\theta d\theta d\phi$. These substitutions lead directly to the evaluation of the solid angle integral, which is discussed first:

$$\Omega_A = \iint_A \sin\theta \, d\theta d\varphi = \sum_{i=1}^{N} \alpha_i - (N-2)\pi - \oint_{\partial A} \kappa_g dl \quad (5)$$

The surface of integration, A, for the solid angle is that shown in Fig. 1. The boundaries of the surface A are arc segments of great circles around the sphere, i.e. circles that circumscribe the diameter of the sphere. For a circle on a unit sphere, the geodesic curvature is given by:

$$\kappa_g = \frac{\sqrt{1-r^2}}{r} \quad (6)$$

Great circles are geodesics with radius r=1 and therefore have geodesic curvature of zero. The equation for solid angle then is reduced to the summation over interior angles, minus the sum of interior angle of the corresponding planar polygon:

$$\Omega_A = \iint_A \sin\theta\, d\theta d\varphi = \sum_{i=1}^{N} \alpha_i - (N-2)\pi \quad (7)$$

The right hand side of this equation is also known as the spherical excess and is a well-known relationship for solid angle that may be used evaluate the integral on arbitrary polygonal sections.

In order to apply the Gauss-Bonnet theorem to the projected solid angle integral, the double angle formula is applied to remove the cosine term:

$$\bar{\Omega}_A = \tfrac{1}{2}\iint_A \sin 2\theta\, d\theta d\varphi \quad (8)$$

The variables of integration are next changed using $\bar{\theta}= 2\theta$ to transform the integral to an equivalent integral over a patch on the unit sphere. This transformed integral can then be solved using Gauss-Bonnet:

$$\bar{\Omega}_A = \tfrac{1}{4}\iint_{\bar{A}} \sin\bar{\theta}\, d\bar{\theta} d\varphi \quad (9)$$

This transformation of variables produces a new surface of integration ($\bar{A}$) by stretching the $\theta$ co-ordinate while leaving the $\phi$ co-ordinate intact. This alters the size of the spherical patch, moving the vertices from ($\theta_i,\phi_i$) to ($2\theta_i, \phi_i$), and also modifies the edges such that their curves are described by $\bar{\theta}(\phi) = 2\theta(\phi)$. The resulting curves, being lines on a spherical surface, are still circles. Thus, the edges are transformed to segments of small circles (i.e. circles that do not circumscribe the full diameter of the sphere). One noteworthy mapping is points on the equator ($\pi/2$) to the south pole ($\pi$). Since every great circle crosses the equator, every corresponding small circle will pass through the south pole. The two vertices and the south pole then completely define each small circle. Note that, for a projected solid angle, the surface patch A is confined to the northern hemisphere ($\theta \leq \pi/2$), so there are no points in $\bar{A}$ that have $\bar{\theta}$ larger than $\pi$. This transformed geometry is shown in Fig. 2(a).

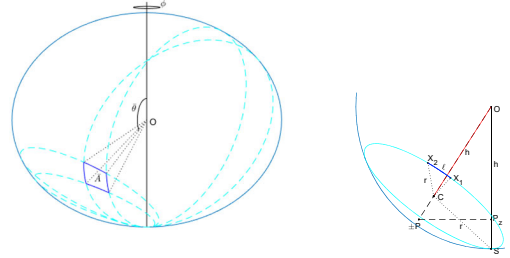

Figure 2. (a) Transformed co-ordinates on unit sphere for evaluation of projected solid angle. (b) Dimension on one edge and corresponding small circle.

It is found numerically that the spherical excess (i.e. the right hand side of (7)) for this surface is zero, regardless of where the patch is located or how many sides it has. This observation for a closed polygon whose circles pass through a common point can likely be proved mathematically, but such a proof is not attempted here. With this observation, the Gauss-Bonnet theorem now reduces the areal integral to the integral of geodesic curvature around its border.

$$\bar{\Omega}_A = -\tfrac{1}{4}\oint_{\partial \bar{A}} \kappa_g dl \quad (10)$$

Each edge has a constant curvature, so that the integral may be rewritten as a summation over edges. Figure 2(b) shows one such edge and its corresponding small circle with dimensions r and l labeled. The center of the small circle (C) forms a right triangle with the center (O) and south pole (S) of the unit sphere. The altitude of the small circle, h, is given by the third leg of the triangle and is equal to $\sqrt{1-r^2}$. With these substitutions the projected solid angle is now given by:

$$\bar{\Omega}_A = -\tfrac{1}{4}\sum_i h_i \frac{l_i}{r_i} \quad (11)$$

The altitude of the circle may be found by first computing the pole that passes through its center. The polar vector is denoted $\pm P$ on Fig.2(b). The ambiguity in sign represents that the pole may be pointing up or down and will be addressed below. The pole is found by taking the cross-product of two vectors in the small circle, then normalizing the result:

$$\vec{P}_i = \frac{(\vec{X}_{i,1}-\vec{S})\times(\vec{X}_{i,2}-\vec{S})}{\|(\vec{X}_{i,1}-\vec{S})\times(\vec{X}_{i,2}-\vec{S})\|} \quad (12)$$

The z-component of $\vec{P}$, given by its projection on the south pole, is denoted as $P_z$ on Figure 2(b). The coordinates $POP_z$ form a congruent triangle to SOC. The distance of segment $OP_z$ is therefore equal to that of OC, which is the altitude. The altitude then is given by $P_z$ and the center is located at $h\vec{P}$. The ratio of the arclength to radius, $l/r$, is the angle subtended from C to $X_1$ and $X_2$ and can be found by taking the arccosine of the vector dot product. These three relationships are summarized below:

$$h_i = -\vec{P}_i \cdot \hat{z}$$

$$\vec{C}_i = h_i \vec{P}_i \quad (13)$$

$$\frac{l_i}{r_i} = cos^{-1}\left(\frac{(\vec{X}_{i,1}-\vec{C}_i)\cdot(\vec{X}_{i,2}-\vec{C}_i)}{\|\vec{X}_{i,1}-\vec{C}_i\|\|\vec{X}_{i,2}-\vec{C}_i\|}\right)$$

The projected solid angle is now found by evaluating eq. (11), substituting Eq (12)-(13) for the quantities therein. It is noted here that the arclength integral in Gauss-Bonnet is signed. The sign depends on the direction taken around $\partial \bar{A}$. While technically the sign should appear on $l_i$, the equations above are written with the sign appearing on $h_i$. (The sign is introduced by the cross product in equation (12) and carries through to $h_i$ in (13)). The sign reverses as the summation traverses the polygon. The magnitude of the projected solid angle is then given by the differences in the line integral of geodesic curvature (equivalently arclength-altitude product) on opposing sides of the polygon. To obtain the correct sign with the above equations, a right-handed rotation about the outward normal should be followed. However, since the projected solid angle must be positive, the absolute value of the summation in either direction may be taken, so long as the vertices are traversed in order.

While the above equations give the complete solution to the projected solid angle, a few notes for calculation are warranted. First, any edge of A on the equator maps to a single point (the south pole) in $\bar{A}$. This creates a singularity where both the arclength and radius of the small circle are zero. The ratio of $l/r$ however is unchanged by this transformation and simply equals $\phi_1$-$\phi_2$. An edge that has a vertex on the equator in A presents a similar problem in evaluating equation (12) and possibly (13). Both of these issues may be resolved numerically by introducing a small positive offset in θ, so that $(\vec{X}_i - \vec{S})$ is approximated by $(\delta \cos\varphi, \delta \sin\varphi, \frac{\delta^2}{2})$ and the factor of δ cancels out in the evaluation. Another issue may occur if the angle around the circle from $X_{i,1}$ to $X_{i,2}$ is larger than π, such that the arccosine returns the length around the wrong side of the small circle. This condition is corrected by subtracting the result from 2π to obtain the correct value. Finally, numerical precision may impact the evaluation of the arclength when the X's are close together such that the dot product is near unity. In this case the distance of the curved line may be approximated by taking the linear separation between the two points on the sphere, i.e. $\frac{l_i}{r_i} \approx \frac{(\vec{X}_{i,1}-\vec{C}_i)}{\|\vec{X}_{i,1}-\vec{C}_i\|} - \frac{(\vec{X}_{i,2}-\vec{C}_i)}{\|\vec{X}_{i,2}-\vec{C}_i\|}$. (Note that, while the magnitude of both vectors is r, the normalization is computed explicitly to minimize numerical error)

A test of the evaluation of the integral is shown in Figure 3, where the projected solid angle is calculated on a hemisphere by summing over surface patches computed numerically and with the above formulae. Fig 3(a) shows the result for patches spaced evenly in θ and ϕ. The numerical integral uses a vertex averaged value of cos θ and sin θ. Fig 3(b) shows an integration performed with triangular surface patches. This numerical integral does not have constant boundaries in θ or ϕ so is converted to cartesian coordinates and the vertex averaged value of cos θ is the integrand. The integral of projected solid angle on a hemisphere is known to equal π, which can be found analytically from Eq. (1) or (9). The numerical integration approaches this limit from below and requires 100 quadrilaterals or 300 triangles to come within 1% of the exact value. The numerical error decreases approximately linearly with the number of elements employed. In contrast, the analytical formulae presented here yield this value within machine precision regardless of the number of triangles or quadrilateral elements employed. Extension to higher order polygons is possible, but not attempted due to the complexity of constructing the corresponding surface mesh. These formulae (11)-(13) may now be employed for calculating the projected solid angle and corresponding view factors to, or irradiance from, free space for any geometry that may be approximated by a number of discrete polygons.

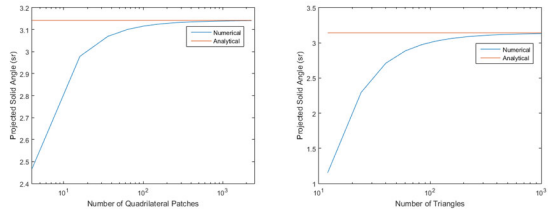

Figure 3. Comparison of numerical integration versus analytical formulae versus number of surface patches employed for (a) evenly spaced quadrilaterals in θ and φ, and (b) triangular surface patch elements